\documentclass[doublecol,linenumbers]{epl2}
\usepackage{amsmath}
\usepackage{amssymb}
\usepackage{amsfonts}
\usepackage{graphicx}
\usepackage{dcolumn}
\usepackage{bm}
\usepackage{sidecap}
\usepackage{float}
\usepackage{parskip}
\usepackage{grffile}

\usepackage{color}
\usepackage{hyperref}
\usepackage{hhline}
\usepackage{mathtools}
\usepackage{slashbox}
\usepackage{epstopdf}
\makeatletter
\def\@eqnnum{{\normalsize \normalcolor (\theequation)}}
 \makeatother
\title{Birth and Death of Chimera: Interplay of Delay and Multiplexing}

\author{Saptarshi Ghosh\inst{1} \and Anil Kumar\inst{1} \and Anna Zakharova\inst{2} \and Sarika Jalan\footnote{Corresponding Author (Email: sarikajalan9@gmail.com)}\inst{1,3}}

\shortauthor{S. Ghosh \etal}

\institute{                    
  \inst{1} Complex Systems Lab, Discipline of Physics, Indian Institute of Technology Indore, Khandwa Road, Simrol, Indore-453552, India \\
  \inst{2} Institut f{\"u}r Theoretische Physik, Technische Universit\"at Berlin, Hardenbergstra\ss{}e 36, 10623 Berlin, Germany \\
  \inst{3} Centre for Biosciences and Biomedical Engineering, Indian Institute of Technology Indore, Khandwa Road, Simrol, Indore-453552, India \\
}


\abstract{{The chimera state with co-existing coherent-incoherent dynamics has recently attracted a lot of attention due to its wide applicability.} We investigate non-locally coupled identical chaotic maps with delayed interactions in the multiplex {network} framework{ and find that an interplay of delay and multiplexing brings about an enhanced or suppressed appearance of chimera state depending on the distribution as well as parity of delay values in the layers.} Additionally, we report a {\it layer chimera state} {with an existence of one layer displaying coherent and another layer demonstrating incoherent dynamical evolution.} The rich variety of {dynamical} behavior {demonstrated} here can be used to gain further insight into the real-world networks which inherently possess {such} multi-layer architecture with delayed interactions.}

\begin{document}

\maketitle


{\bf Introduction.} Over {the} last few decades, the field of network science has witnessed tremendous growth {owing to} its applicability in real-world systems ranging from biology and sociology to economics {\cite{barabasi_rev,socio_Plos_one,bio_sci_rep}}. One of the prime focus {of network science is} to study the collective {behaviour} of dynamical units {coupled on these networks \cite{dynm_1}}. {Coupled dynamical units are shown to exhibit} a plethora of novel phenomena \cite{dynm_2}. One such phenomenon is {the emergence of} chimera \cite{chim_review}. Initiated by {the} works of Kuramoto, Battogtokh and Shima {\cite{chim_def_1}} and later christened by Abrams and Strogatz \cite{chim_def_3}, chimera state represents a hybrid dynamical state with co-existing coherence and incoherence in an identical network with symmetric coupling environment. Since then, numerous theoretical \cite{theo_chim_1,theo_chim_2} as well as experimental investigations \cite{exp_chim_1,exp_chim_2} have been reported for the observation of chimera state in various systems \cite{chim_review,osc_chim_2,theo_chim_2,exp_chim_2}. Moreover, different types of chimera state have been proposed to incorporate a variety of stable {and} unstable co-existing coherent-incoherent state \cite{chim_review}. Chimera states {have} provided a framework {to understand} epileptic seizures \cite{epileptic_1,epileptic_2} {and} unihemispheric sleep recently observed in human \cite{TAM16}. It has {also} been associated with motion of heart vessels to explain ventricular filtration \cite{ventricular} as well as with the so-called bump state in neural networks \cite{bump}.

{Furthermore,} recent years have witnessed a burst of studies pertaining to the analysis of multiplex networks which is defined as a collection of two or more layers having mirror nodes in each layer \cite{mul_def}. {An} objective of multiplex network framework is to study multiple levels of interactions where function of one layer gets affected by the properties of {other layers} \cite{mul_review_1}. 

Yet another approach for understanding real world systems through network theory is to incorporate time delay{s}. {An} analysis of impact of time delay on dynamical properties of a coupled system is very crucial {to predict and explain dynamic evolution of such systems} \cite{delay_def}. {Numerous dynamical phenomena}, including enhancement or suppression of synchronization, have been {found for} time delayed networks \cite{delay_appl}. Although, it has been recently shown that chimera states exist in time delayed networks \cite{theo_chim_1,delay_chim_expl}, the occurrence of {this} complex spatio-temporal pattern is yet not well understood in the presence of time delay {in multiplex networks}.

We investigate {an} interplay of the delay and the multiplexing on the occurrence of chimera state. Delay coupled systems are known to exhibit globally clustered chimera state \cite{delay_chim}. However, we find that small delay{s lead to an} enhance{ment} or suppress{ion in} the chimera depending upon the parity of the delay, {and} large delays always lead to the suppression in the chimera state. In multiplex networks, this enhancement or suppression depends upon the distribution of delay in the individual layers which further results in a new type of chimera state, henceforth called as {\it layer chimera state}. This pattern exhibits {a} co-existence of coherent and incoherent dynamics in different layers which are unique to the delayed multiplex systems. We investigate the emergence of the chimera state and its dependence on the delays and various other properties of the layers {of a} multiplex network.

\begin{figure}[t]
 \centerline{\includegraphics[width=2.5in, height=1.0in]{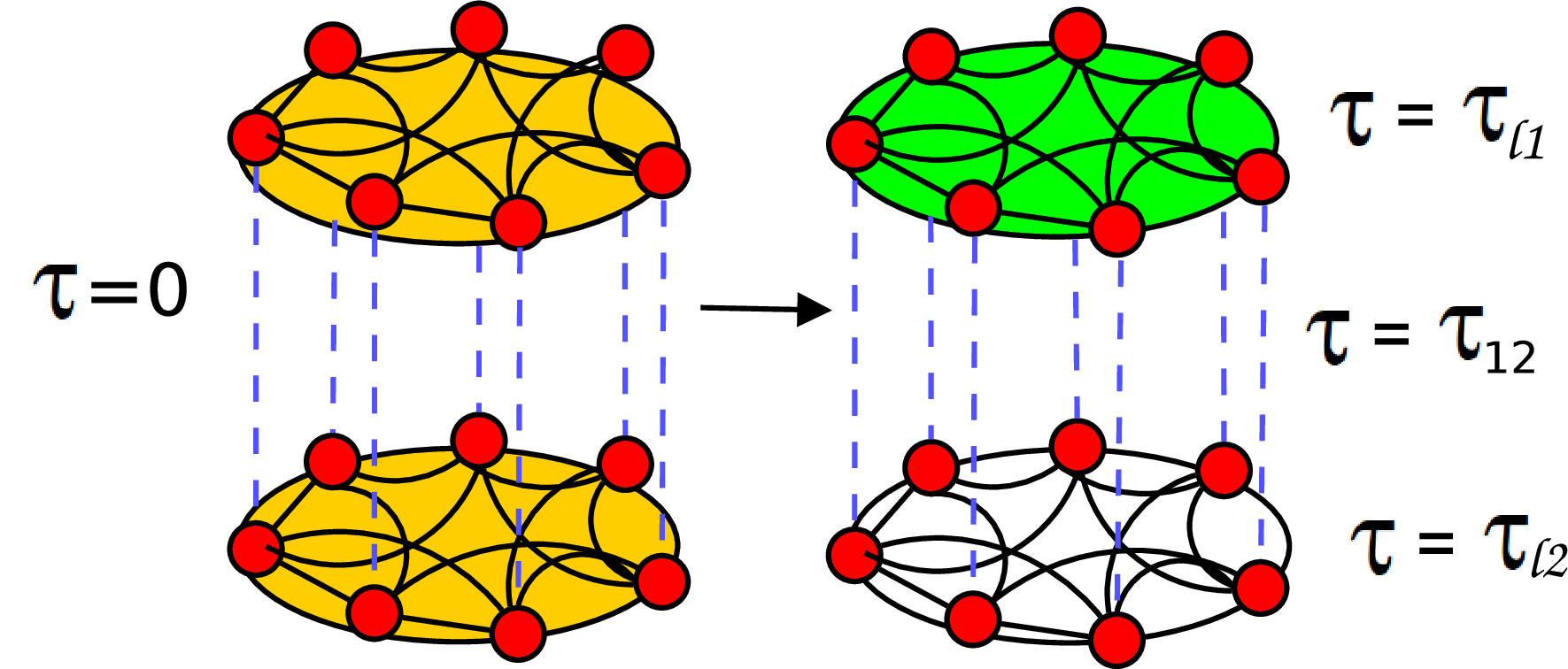}}
 \caption{ {(Color online)} Schematic diagram for multiplex network consisting of two layers. 
 Each layer is represented by identical 1D lattice with non-local interactions. 
 Each node (circle) has the same coupling architecture. {$\tau_{l1}$($\tau_{l2}$)} signifies delay in intra-layer connections (represented as solid lines) and $\tau_{12}$ depicts delay in the inter layer connections of the multiplex network (represented as dashed lines).} 
 \label{multiplex} 
 \end{figure}

{\bf Model.} {Let us} consider a multiplex network of $2N$ nodes where each layer is represented by {a} 1D ring lattice consisting of $N$ nodes (Fig.~\ref{multiplex}).
The dynamical evolution of each {node} is represented by $z_{i}(t)$ where $i=1,2,...,2N$. This evolution can be written incorporating the time delay $\tau$ as 
\begin{equation}
z_{i}(t+1)=(1-\varepsilon) f(z_i(t))+\frac{\varepsilon}{k_i} \sum_{j=1}^{2N} A_{ij} [ f(z_i(t-\tau)) ],
\label{eq.evol}
\end{equation}

where $\varepsilon$ represents {the} overall coupling strength and $k_i=(\sum_{j=1}^{2N}A^l_{ij})+1$ is the degree of the $i^{th}$ node. {$\tau$} introduce delay values depending on the multiplex adjacency matrix as illustrated in Fig.~\ref{multiplex}. The function $f(x)$ is a non-linear {logistic Map given as $f(x)= \mu x (1-x) $ \cite{logsMap_def}}.

{Coupled logistic maps display a rich dynamical behaviour depending on the values of $\mu$. Due to its simplicity, yet ability to display complex chaotic behavior, coupled logistic maps have been widely investigated to understand various complex phenomena manifested by a diverse range of real-world systems \cite{dynm_1}.}

{We use $\mu=3.8$ for which $f(x)$ shows {the} chaotic behaviour \cite{logsMap}.} We consider the coupling {radius} $r=0.32$, {where $r$ represents the coupling radius defined by r = P/N, with P signifying the number of neighbors in each direction in a layer.} The adjacency matrix ($A$) of {a} multiplex network with two layers can be written as


\begin{equation}
   A=
      \begin{pmatrix} 
      A^1 & I  \\ 
      I & A^2  \\
      \end{pmatrix} 
\end{equation}

For the adjacency matrix of the $k^{th}$ layer ($A^{k}$), each element is defined as $A^k_{ij}=1 (0)$ depending upon whether $i^{th}$ and $j^{th}$ nodes are connected (or not) in the $k^{th}$ layer. The matrix $A$ is {a} symmetric matrix with diagonal entries $A_{ii}=0$ depicting no self-connection. $I$ is a unit $N$x$N$ matrix representing one to one connection between the mirror nodes.{ We consider bi-directional inter-layer connections which maintain symmetric coupling environment required for defining a chimera.}

A {spatially coherent} state of a network is {considered} as, if for any node $i$ belonging to a particular layer, the spatial distance between the {amplitude of} neighboring nodes approaches to zero for $t \rightarrow \infty$,
{
\begin{equation}
\lim\limits_{t \rightarrow \infty} \mid{z_{i+1}(t) - z_i(t)}\mid \rightarrow 0, \, \, \, \, \forall \; i \in S^1
\label{eq.cohr}
\end{equation}
} whereas, temporal coherence {is defined} as, {$
\lim\limits_{t \rightarrow \infty} \mid{z_{j}(t) - z_i(t)}\mid \rightarrow 0$ for $\forall \; i,j \in S^1$ \cite{theo_chim_2}.} Therefore, temporal coherence can be written as {$z_1(t)=z_2(t)=\cdots= z(t)$} which leads to a straight line for the spatial curve { in $z_i(t)-i$ plane}. This straight line represents a completely synchronized state {($z_i(t)=z_j(t) \forall i,j$)}. Now due to {a} near zero local spatial distance between {the} neighboring nodes, the coherent state can be geometrically represented by a smooth curve {\cite{theo_chim_2}}. Any discontinuity appearing in the curve represents co-existence of coherence-incoherent state leading to the chimera state.

\vspace{3mm}

\begin{figure}[t]
 \centerline{\includegraphics[width=1\columnwidth]{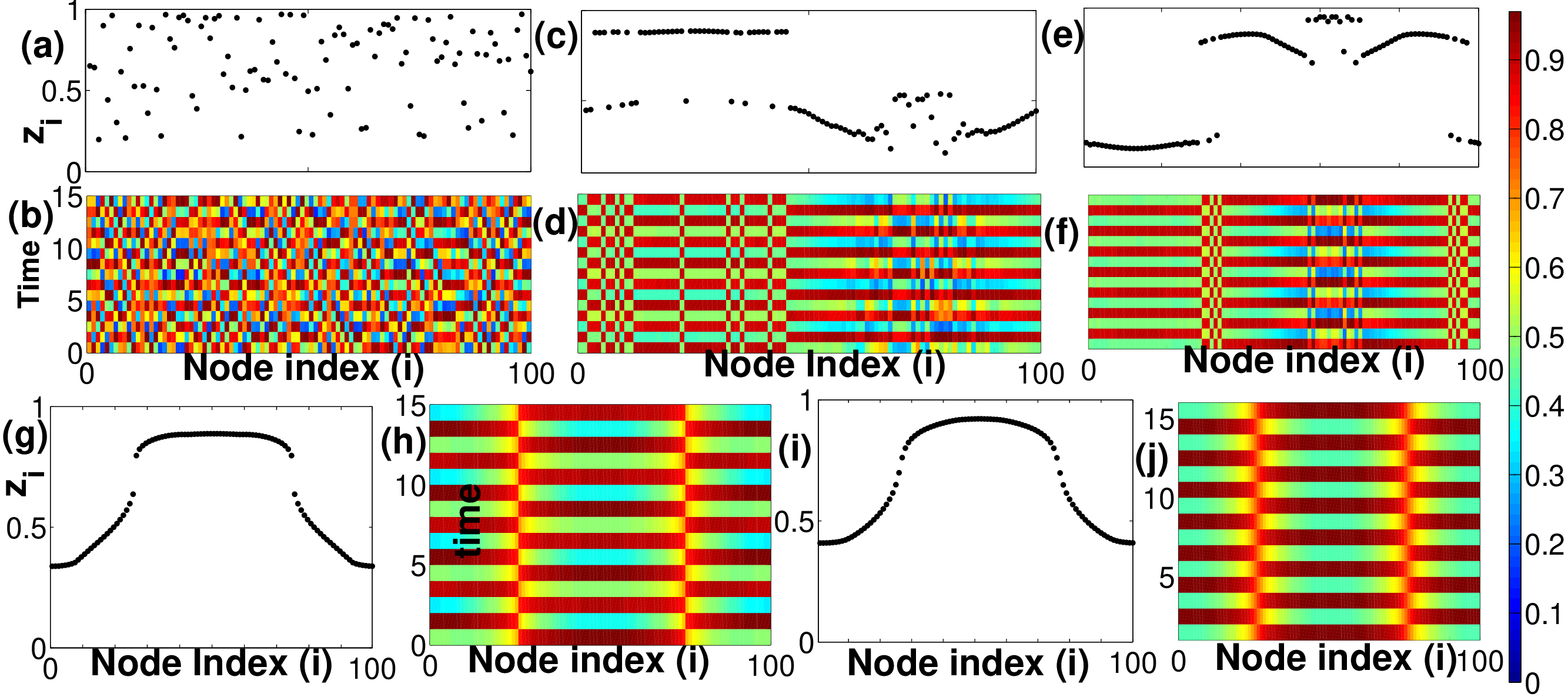}}
 \caption{ {(Color online)} Snapshots and space-time plots for isolated ring network for different coupling strengths (a) and (b) $\varepsilon=0.1$, (c) and (d) $\varepsilon=0.32$, (e) and (f)  $\varepsilon=0.36$, (g) and (h) $\varepsilon=0.46$,(i) and (j) $\varepsilon=0.5$. Dynamical variable associated with the evolution is color-coded in the space time plots as a function of time. Other network parameters are $N=100$ and $r=0.32$, $\tau=0$.} 
 \label{iso1}
 \end{figure}
We quantify the absence of smoothness by a distance measure based on the spatial distance which can be written as 
{
\begin{equation}
d_{i}(t)=| (z_{i+1}(t)-z_{i}(t))-(z_{i}(t)-z_{i-1}(t))|
\label{eq.dmsr}
\end{equation}}
Due to a near zero spatial distance between the neighboring nodes, 
the coherent state can be geometrically represented by a smooth spatio-temporal curve
plotting  amplitudes of all the nodes as a function of node indices for a particular time 
value. Since, the chimera state is represented by a co-existence of the coherent-incoherent dynamical evolution of the nodes, the spatio-temporal plot consisting of a smooth part (continuous curve representing coherent nodes) and gaps (discontinuities representing incoherent nodes ) reflect an existence of the chimera state.
This measure capturing the difference of the spatial distance between the neighboring nodes,{ attains a value towards} zero for {a} coherent profile. Any discontinuity in the spatial curve is indicated by a kink in {$d_i$-i plane \cite{theo_chim_2}}. We use this measure to find the chimera states as discussed in the next section.
\begin{figure}[t]
 \centerline{\includegraphics[width=3.0in, height=1.5in]{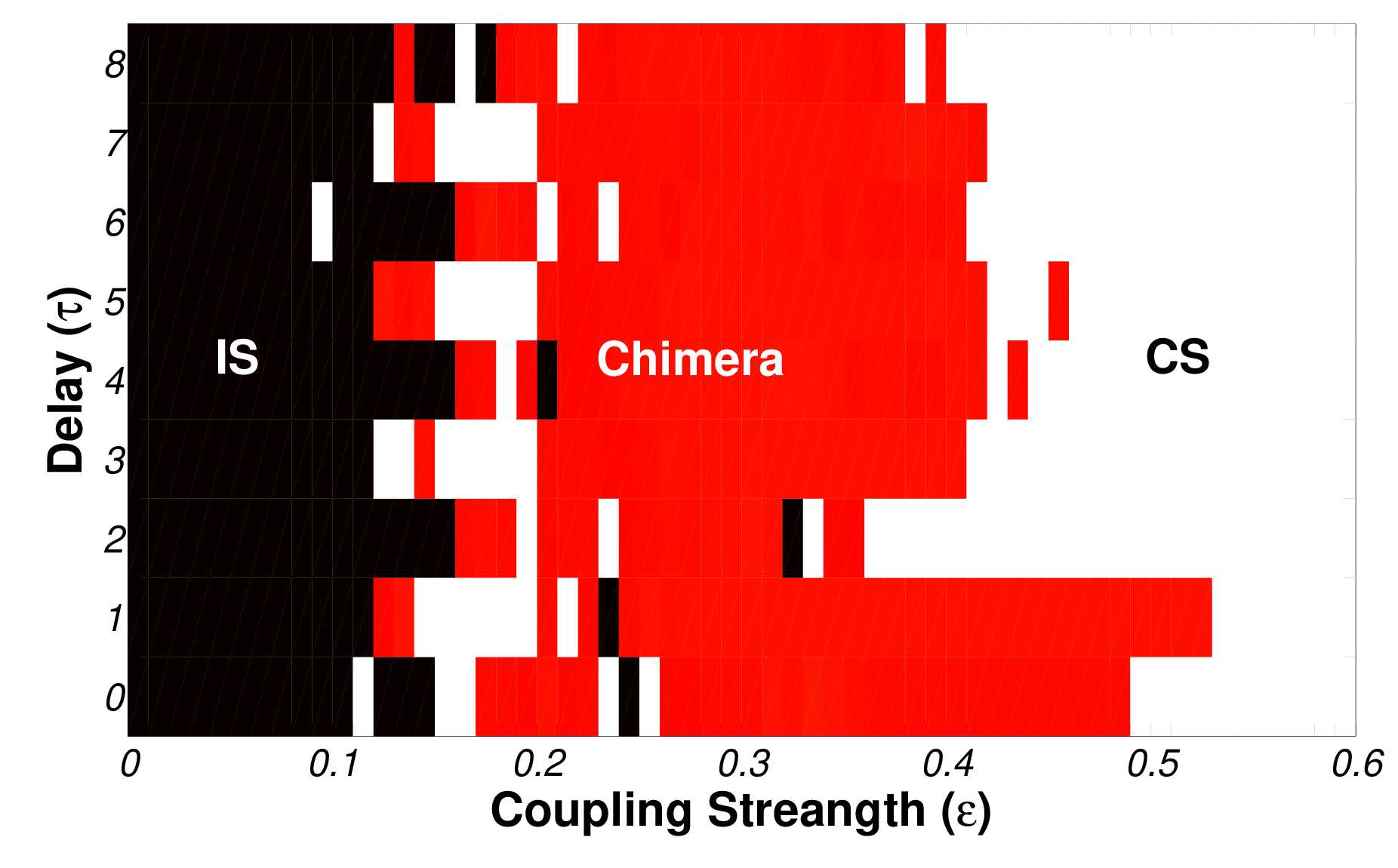}}
 \caption{ {(Color online)} Phase diagram showing different dynamical regions based on number of spatial clusters ($N_{clus}$) in parameter space of delay ($\tau$) and coupling strength ($\varepsilon$) for isolated ring network. The {shades} (colors) denotes different regions: IS (incoherent state {$N_{clus}=0$}), CS (coherent state {$N_{clus}=1$}) and chimera (chimera state {$N_{clus}>1$}). Other network parameters taken are $N=100$ and $r=0.32$.}
 \label{iso2}
 \end{figure}
 \begin{figure}[t]
 \centerline{\includegraphics[width=2.8in, height=1.5in]{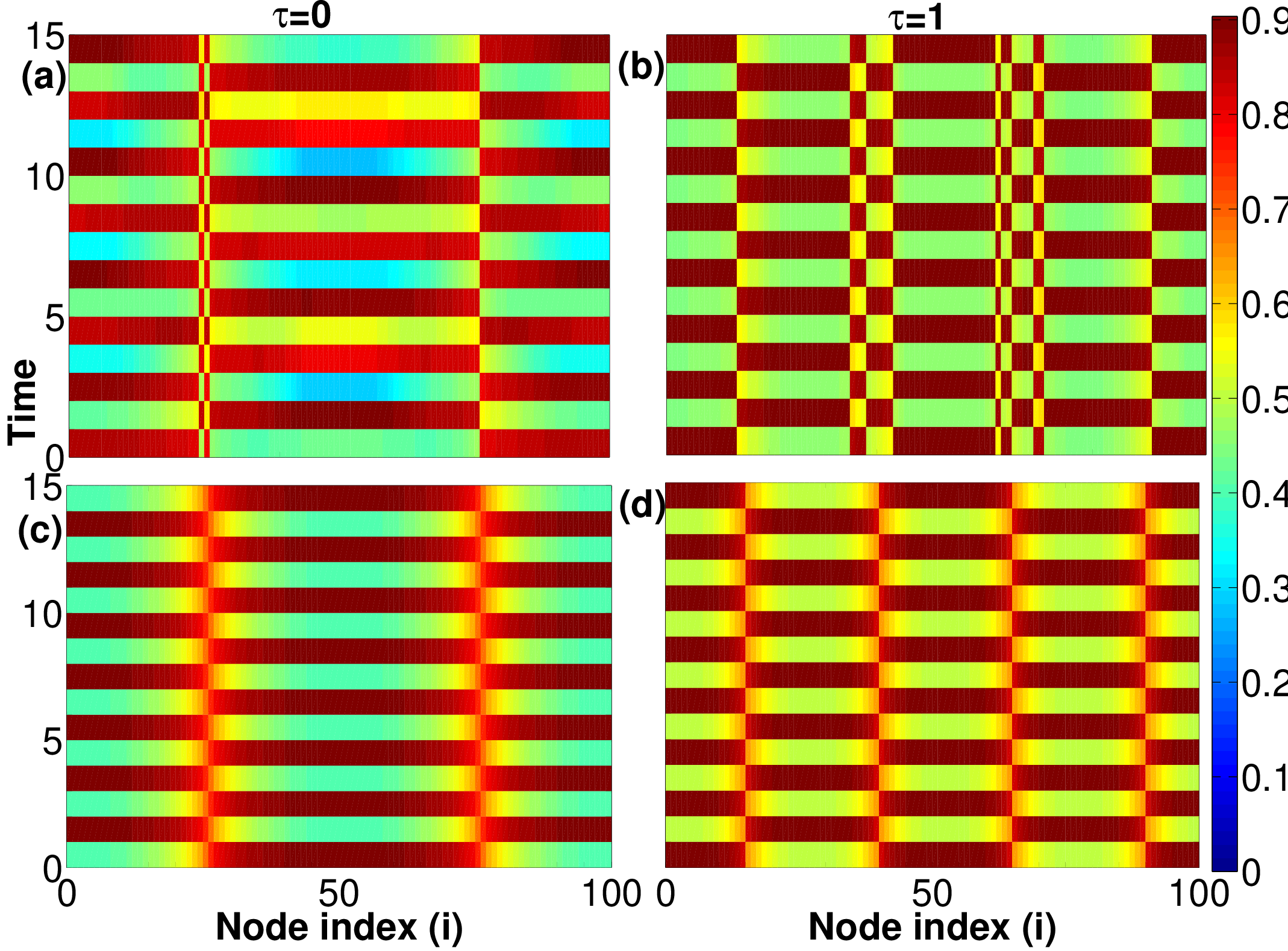}}
 \caption{ {(Color online)} Space-time pattern for the chimera state of the (a) and (c) undelayed ($\tau=0$)  and (b) and (d) delayed ($\tau=1$)  evolution for coupling strengths $\varepsilon$=0.39 (a) and (b)  and $\varepsilon$=0.49 (c) and (d) in isolated ring network. Dynamical variable associated with the evolution is color-coded in the space time plots as a function of time. Other network parameters are $N=100$ and $r=0.32$.} 
 \label{iso3}
 \end{figure}

{\bf Initial conditions and numerical considerations.} The chimera state has been reported to be sensitive to the {choice of} initial conditions. A proper choice of the initial condition is required for the dynamical state of a system to show either the chimera state or the complete coherent state \cite{init_condtn}. {For example, a special humpback function was used to define initial condition to demonstrate occurrence of chimera state in identical non-locally coupled kuramoto oscillators \cite{chim_def_3}.} However, few recent works have suggested that the {choice of} initial condition for {occurrence of} the chimera state may not be so significant. {For instance,} Ref.~\cite{exp_chim_1} has shown the chimera-like behavior with {a set of} random initial conditions. Ref.~\cite{osc_chim_2} utilized a quasi-random initial condition for the realization of the chimera state in {a} coupled photosensitive chemical oscillators. Here, we consider a uniform random distribution of initial states ($z_0(i)$) for the $i^{th}$ oscillator which is bound between an interval $[ 0 , exp\big[-\frac{(i-\frac{N}{2})^2}{2\sigma^2}\big])$ where the variance $\sigma$ is chosen depending on the size of the network such that $z_0(i)\in [0 , 1]$. We use the same set of initial conditions for both the layers in a multiplex network {and} find that despite choosing a quasi-random initial condition, {introduction of delay in a layer leads to an enhanced or a suppressed critical coupling strength for coherent evolution of dynamical variable in that layer}, depending {upon distribution and parity of delay values in different layers.}

Furthermore, {the} number of spatial clusters is used
{based on a collection of nodes with coherent evolution} to identify the appearance of the chimera states. These spatial clusters are counted through the identification of spatially neighboring nodes having distance measure $d_i<\delta$, where $\delta$ is a small quantity. {We choose a small positive threshold value $\delta$ ($\delta \approx 0.0384$) to clearly distinguish different dynamical states \cite{thrshld}.} We discard clusters with the node population {below a certain threshold, for instance 5, as considered in few other works \cite{thrshld}.} The number of spatial clusters identifying different types of dynamical states can be written as follows; $N_{clus}=0$ for the spatially incoherent state, $N_{clus}=1$ for spatially coherent or completely synchronized state and $N_{clus}>1$ for the chimera state.

\vspace{5mm}
{\bf Delayed evolution in an isolated network.} First, we discuss the impact of delay on the dynamics of a single layer ring network. For the undelayed evolution, {the} isolated ring network exhibits a transition from the incoherent to the coherent state via chimera pattern {as coupling strength ($\varepsilon$) is increased}. In the absence of any delay ($\tau=0$), in the weak coupling range, the spatial profile indicates an incoherent dynamics with no correlation between the spatial neighboring nodes. For example, {we find that} as we increase the coupling strength from $\varepsilon=0.1$, an emergence of the partial coherent states is observed (Fig.~\ref{iso1}(a) and Fig.~\ref{iso1}(b)). For {mid-range coupling} ($0.16\le \varepsilon \le 0.49 $), the chimera state can be seen (Figs.~\ref{iso1}(c), \ref{iso1}(h)). Further increment in the coupling strength leads to a decrement in the incoherent regions. For $\varepsilon=0.46$, the incoherent regions shrink to the point discontinuities in the otherwise {a} spatial smooth profile of the coherent nodes and we observe two distinct spatial clusters (Fig.~\ref{iso1}(g), \ref{iso1}(h)). For strong coupling strength ($\varepsilon \ge 0.5$), the dynamical evolution of {the nodes exhibiting a} chimera state revert to {a} completely coherent state as {depicted} in Fig.~\ref{iso1}(i), \ref{iso1}(j).{ Fig.~\ref{iso2} displays {a} transition from {the} chimera to {the} coherent state at $\tau=0$ (the last row) in terms of the $N_{clus}$ employed to distinguish between different dynamical states.}

An interesting phenomenon is {discerned} for the mid-range coupling values when the delay is introduced in the isolated ring network. The overall dynamical behavior, as a function of coupling strength, remains same as for the small delayed evolution (Fig.~\ref{iso2}). For the weak coupling strength, the delayed evolution also leads to the incoherent dynamics as found for the undelayed case. For mid-range coupling values, chimera dynamics is observed but the transition from {the} chimera to the coherent state becomes highly dependent on the parity of the delay values (Fig.~\ref{iso2}). We find that {the} chimera state is enhanced for the odd delay {and} exists for {a} larger range of the coupling strength (Fig.~\ref{iso2}), {thus} {being} characterized by {a} high critical coupling strength for the transition from the chimera state to the coherent state. Interestingly, as we increase the {value of} delay, we observe an immediate suppression of the chimera state leading to a {smaller} critical coupling strength (Fig.~\ref{iso2}). {Furthermore,} for the smaller delays, chimera state is found to be enhanced or suppressed depending upon the parity of the delay {value}. For example, we observe critical coupling strength $\varepsilon_{critical}=0.54$ {and} $0.37$ for delay $\tau$=1 and $\tau$=2, respectively as displayed in Fig.~\ref{iso2}. However, as we increase the delay, the chimera state is found to remain suppressed as compared to {the} undelayed case regardless of the parity of delay. {This} suppression of the chimera state becomes dominant for the large delay values. As shown in Fig.~\ref{iso2}, the isolated network shows {a} similar critical coupling strength ($\varepsilon_{critical}\approxeq0.42$) for large delay values ($\tau$=7 and $\tau$=8).

The delayed dynamics shows {a better} spatial clusters {formation} for mid-range coupling strength as exhibited in Fig.~\ref{iso3}. An introduction of the delay enhances {the} incoherence in the chimera state, the incoherent dynamics of the chimera state {becomes} larger and more pronounced. For strong coupling strength, the {dynamical evolution} of the nodes changes from a partially coherent state (incoherent regions shrinked) to {a} completely coherent state (smooth spatial profile) (Fig.~\ref{iso2}).

Furthermore, we find that as {the} time delay increase{s}, the emergence of chimera state becomes suppressed leading to a completely coherent state for {the} mid-range coupling values (Fig.~\ref{iso2}). {This observation is not surprising as} delays {are} known to enhance the synchronization \cite{delay_appl}. {However,} interesting enough, while high delay value enhances the synchronization leading to suppression of chimera dynamics, low delay value is shown to enhance the chimera state (Fig.~\ref{iso2}).
\begin{figure}[t]
 \centerline{\includegraphics[width=2.8in, height=1.5in]{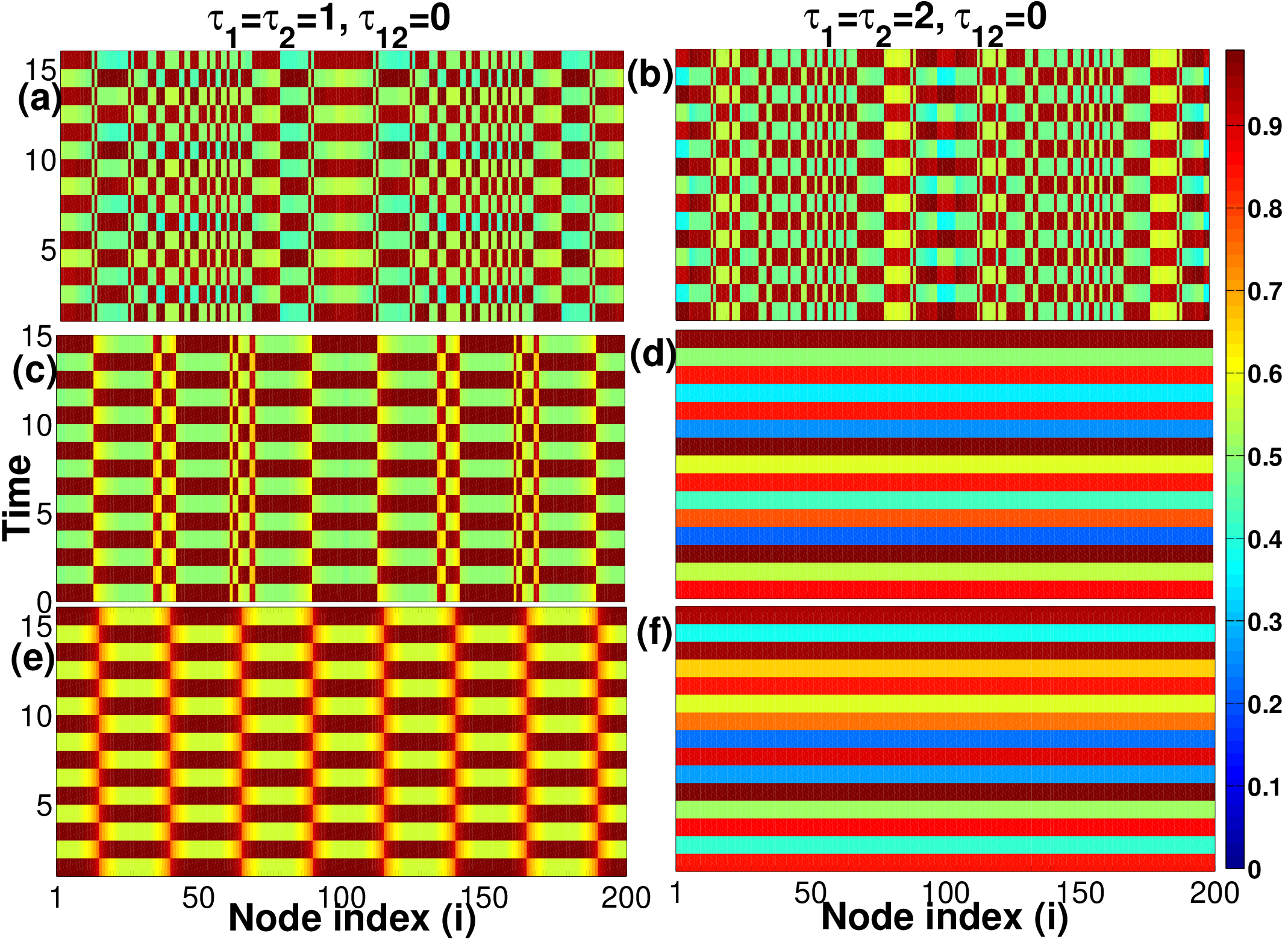}}
 \caption{ {(Color online)} Space-time pattern for the chimera state of (a), (c) and (e) odd delayed {($\tau_{l1}=\tau_{l2}=1$)} and (b), (d) and (f) even delayed {($\tau_{l1}=\tau_{l2}=2$)} multiplex ring network for coupling strengths (a) and (b) $\varepsilon = 0.31$, (c) and (d) $\varepsilon = 0.4$ and (e) and (f) $\varepsilon=0.5$. Intra-layer delay in both layers is the same. The space-time plots presented here are color coded for the first layer (with identical second layer) of the multiplex network. Dynamical variable associated with the evolution is color-coded in the space time plots as a function of time. Other network parameters for each layer of the multiplex network are $N=100$ and $r=0.32$.} 
 \label{mul_1}
 \end{figure}

\vspace{3mm}
{\bf Role of delay in a multiplex network: symmetric intra-layer delay.} Next, we focus to delayed evolution in the multiplex network. Without any delay, the multiplex network exhibits a chimera state for the {mid-range} of coupling values. The dynamical behavior of the individual layer of the multiplex network is found to be exactly same to each other leading to the identical spatio-temporal patterns. {Here}, we consider a multiplex network {with two layers} to present our results. First, {for} introduction of the same delay in both {the} layers of the multiplex network with no inter-layer delay, we find that even with the introduction of delay{s which is symmetric in both the layer,} the multiplex chimera behavior is quite similar to the isolated layers. Initially, {for very small coupling strength value ($\varepsilon \leq 2$)}, we find incoherent dynamics in both the layers for weak coupling range. As the coupling strength {is increased}, {the} chimera state emerges for mid-range coupling values followed by {a} coherent state in both the layers for {the} strong coupling.

Fig.~\ref{mul_1} presents {the} spatio-temporal patterns for different values of {the} delay. The patterns demonstrate a strong similarity with the layers of the multiplex network. Therefore, similar intra-layer delay with no inter-layer delay leads to the similar chimera like behavior {as displayed by the isolated network (Fig.~\ref{iso2})}. However, the nature of the delay plays a critical role in the transition from the chimera to the coherent state. Analogous to the case of an isolated network, we find {that} small odd delay values lead to {an} enhancement of the chimera state with high critical coupling strength ($\varepsilon_{critical}=0.54$ for $\tau=1$), followed by an immediate suppression for small even delay value{s}. The critical coupling strength for intra-layer delay $\tau=2$ is found to be $\varepsilon_{critical}=0.37$. 
\begin{figure}[t]
 \centerline{\includegraphics[width=2.8in, height=2.0in]{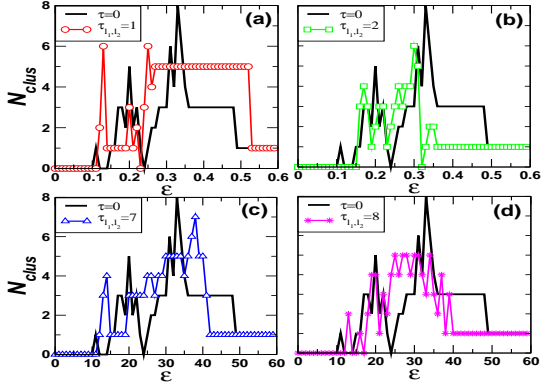}}
 \caption{ {(Color online)} Scatter diagram depicting number of spatial clusters ($N_{clus}$) as a function of coupling strengths ($\varepsilon$) for different delay values (a) {$\tau_{l1}=\tau_{l2}=1$}, (Red circle) (b) {$\tau_{l1}=\tau_{l2}=2$} (Green square), (c) {$\tau_{l1}=\tau_{l2}=7$} (Blue triangle), (d) {$\tau_{l1}=\tau_{l2}=8$} (Magenta star). The thick black line represents $N_{clus}$ for the undelayed case. The values demonstrated here are for first layer (with identical second layer and same intra-layer delay). Other parameters are $N=100$ in each layer and coupling radius $(r)=0.32$.} 
 \label{mul_2}
 \end{figure}
Fig.~\ref{mul_2}(a) and Fig.~\ref{mul_2}(b) display the enhancement and subsequent suppression {of the value of critical coupling strength ($\varepsilon_{critical}$)} as compared to the undelayed case. Depending on the nature of the delay, the chimera state can be enhanced or suppressed to collapse into a coherent state. Moreover, we find that this enhancement is destroyed for high delay values. {A} high intra-layer delay always leads to the suppression of the chimera state as compared to the undelayed case. Fig.~\ref{mul_2}(c) and Fig.~\ref{mul_2}(d) display the range of {coupling strength ($\varepsilon$) for which} chimera state {is observed} for high even and odd delays. {A} vital point to note here is that despite having intra-layer delays, the individual layers of the multiplex network is found to behave like an isolated network as long as the delays in {both} the layers remain the same. 

   \begin{figure}[t]
    \centerline{\includegraphics[width=1\columnwidth]{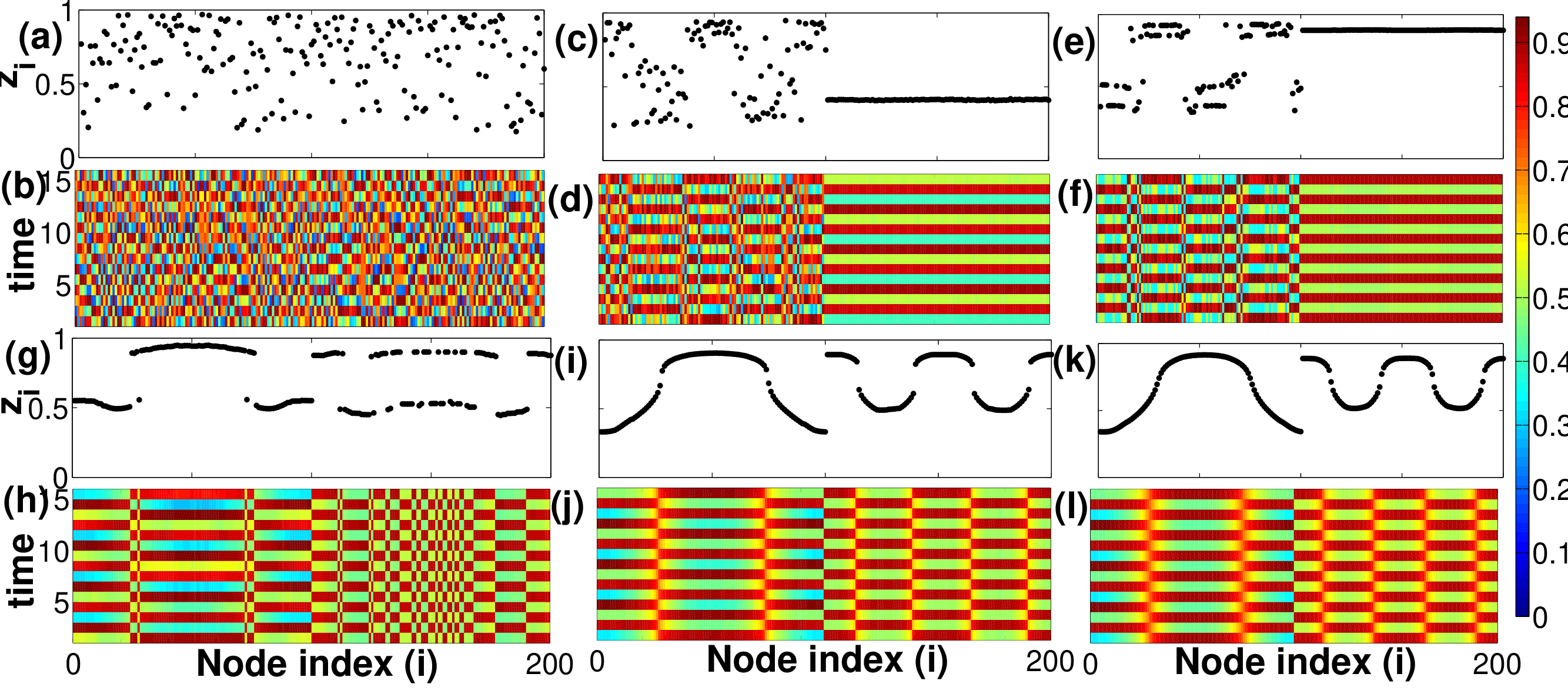}}
    \caption{ {(Color online)} Snapshots and space-time plots for isolated ring network for various coupling strengths (a) and (b) are for $\varepsilon=0.1$, (c) and (d) $\varepsilon=0.14$, (e) and (f) $\varepsilon=0.17$, (g) and (h) $\varepsilon=0.24$, (i) and (j) $\varepsilon=0.49$, (k) and (l) $\varepsilon=0.52$. Dynamical variable associated with the evolution is color-coded in the space time plots as a function of time. Other network parameters for each layer of the multiplex network are $N=100$, $r=0.32$, {$\tau_{l1}=\tau_{12}=0$} and {$\tau_{l2}=1$.}} 
    \label{mul_3}
    \end{figure}

    \begin{figure}[t]
      \centerline{\includegraphics[width=3.0in, height=2.0in]{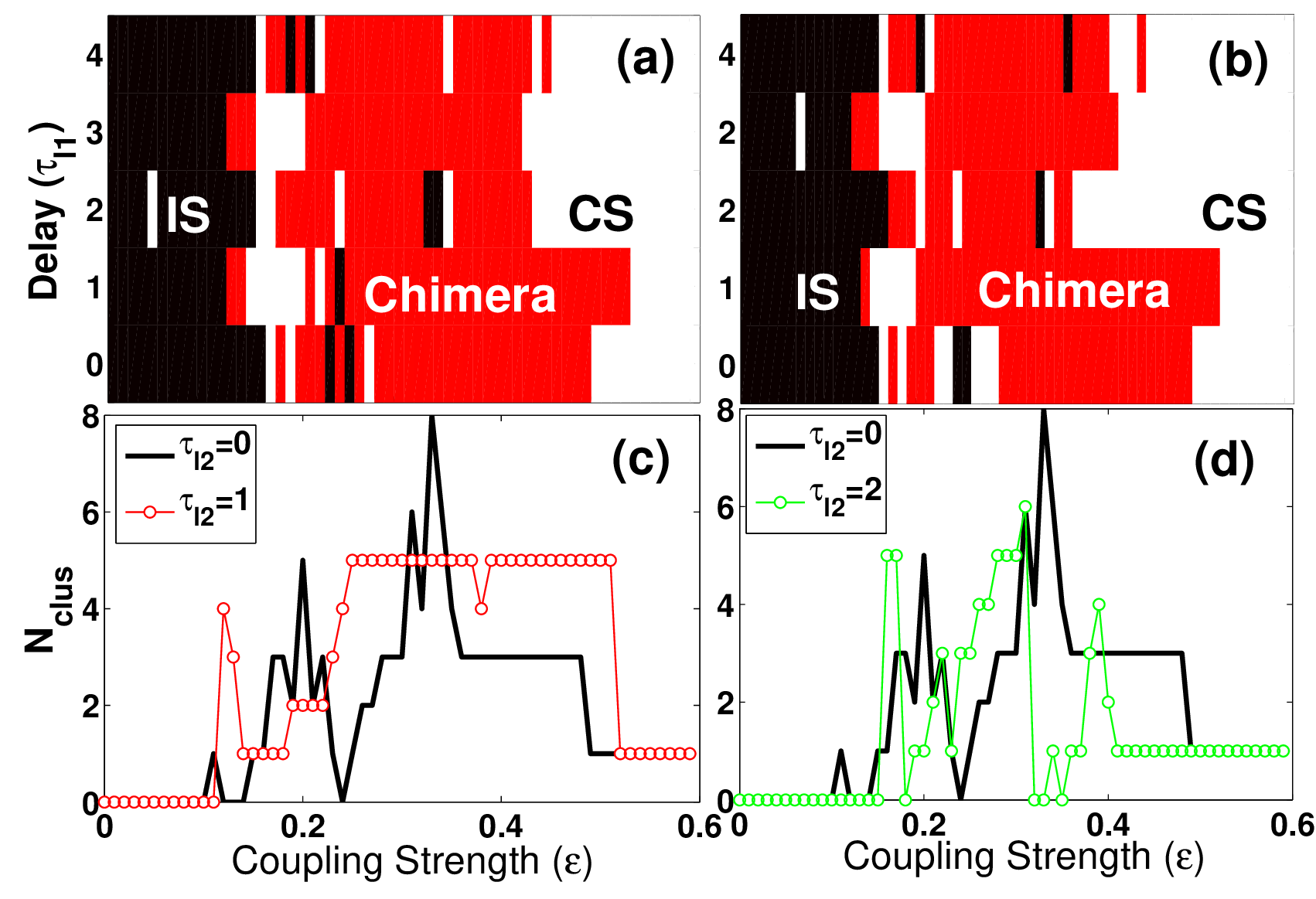}}
      \caption{ {(Color online)} Phase diagram and scatter diagram depicting different dynamical regions for different layers for multiplex network with different intra-layer delays in layers of multiplex network. The {shades} (colors) in phase denotes different regions: IS (incoherent state {$N_{clus}=0$}), CS (coherent state {$N_{clus}=1$}) and chimera (chimera state {$N_{clus}>1$}). (a) and (b) represents $N_{clus}$ values for different values of delay in first layer where as (c) and (d) represents $N_{clus}$ as a function of coupling strength ($\varepsilon$) with constant delay in second layer ($\tau_{l2}=1$ (c)) and $\tau_{l2}=2$ (d)). Other network parameters for each layer of the multiplex network are $N=100$ and $r=0.32$.} 
      \label{mul_4}
      \end{figure}

   \begin{figure}[h]
      \centerline{\includegraphics[width=4.0in, height=1.5in]{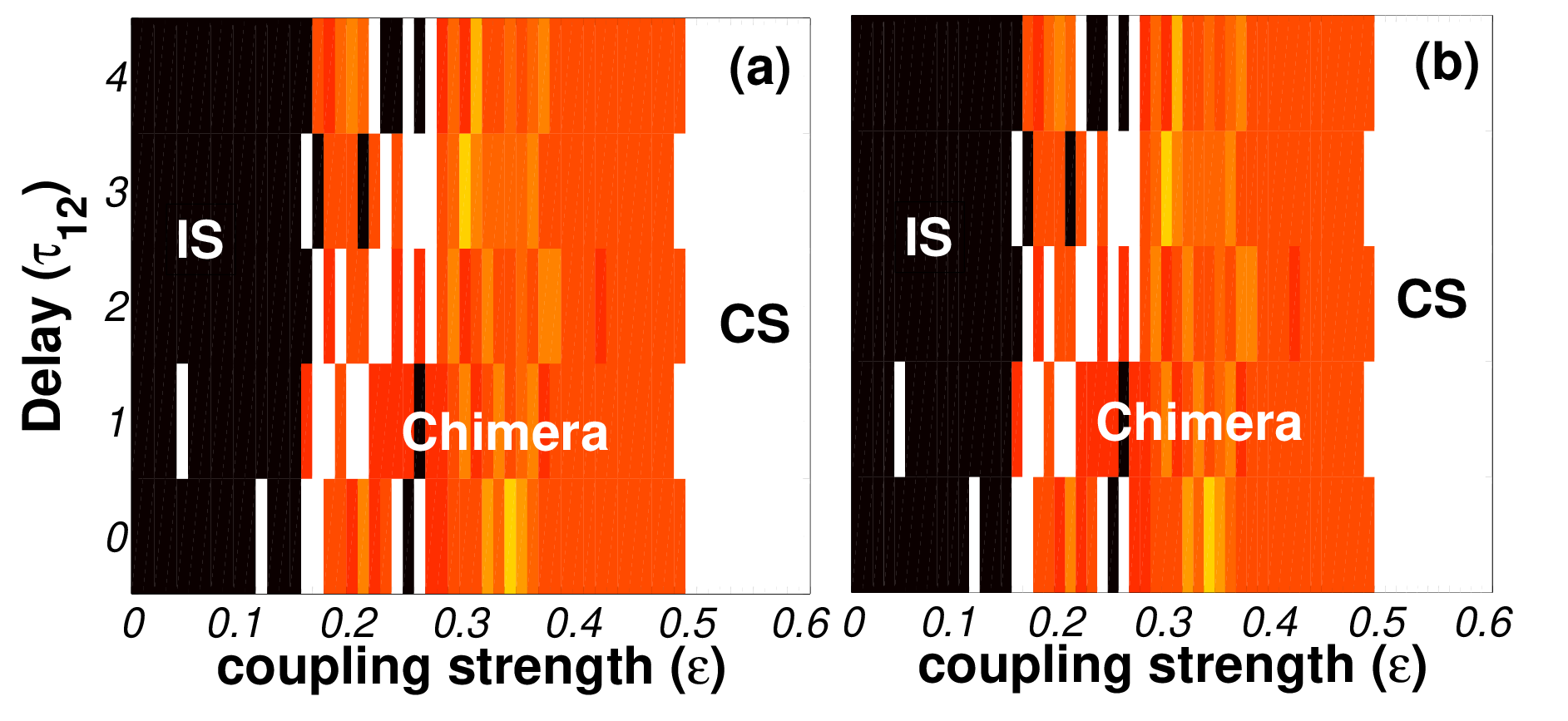}}
      \caption{{ {(Color online)} Phase diagrams depicting different dynamical regions based on number of spatial clusters ($N_{clus}$) in parameter space of inter-layer delay ($\tau_{12}$) and coupling strength ($\varepsilon$) for multiplex ring network. The {shades} (colors) denotes different regions: IS (incoherent state {$N_{clus}=0$}), CS (coherent state {$N_{clus}=1$}) and chimera (chimera state {$N_{clus}>1$}). (a) represents $N_{clus}$ values in first layer where as (b) repreents $N_{clus}$ in second layer with different inter-layer ($\tau_{12}$) delay values. Other network parameters taken are $\tau_{l1}=\tau_{l2}=0$, $N=100$ and $r=0.32$.}} 
      \label{mul_5}
 \end{figure}

{\bf Role of delay in a multiplex network: asymmetric intra-layer delay.} Next, we study the trade off between {the} delay and {the} multiplexing {for occurrence of} the chimera dynamics {for delayed} multiplex network{s}. In real-world systems, having a similar delay {for} all the layers of the {corresponding} multiplex network is very rare. Asymmetric intra-layer delay in layers of the multiplex network leads to a rich variety of emergence of {the} chimera state.
First, we {investigate} the dynamical behavior of the {nodes} when a small {value} of delay is introduced into {only} one layer. The chimera dynamics of the delayed layer is found to be enhanced due to the {introduction of small value of} odd delay {(Fig.~\ref{mul_4})}. Mismatch in the delay value leads to {a} suppression of {the} identical behavior {of the} layers {(Fig.~\ref{mul_3})}. We find that for weak coupling strength, the delay (say in layer 2) enforce{s} a coherent dynamics in the delayed layer while the undelayed layer (say layer 1) still {keeps showing an} incoherent dynamics {(Fig.~\ref{mul_3})}. Although the nodes in one layer are identically connected to {their} mirror nodes in another layer, we witness a surprising co-existence of {the} coherent and {the} incoherent dynamic {evolution of nodes} in the layers of the multiplex network. One population {(layer)} exhibits {a} spatial synchrony while its mirror population retains asynchronous behavior. We term this particular state as {\it layer chimera} state which can only emerge in a delayed multiplex network. With {an} increment in the coupling strength, both layers start exhibiting {the} chimera state. For strong coupling strength, the undelayed layer reaches {to the} coherent state before the delayed layer as {depicted} in Fig.~\ref{mul_3}. 

Now keeping {the} same delay value in one layer ({say} in layer 2), we increase the delay in another layer (in layer 1). The same {value of} delay in both the layers leads to the enhancement in {the} chimera behavior (Fig.~\ref{mul_2}). However, we find that the nature of the chimera dynamics in layer 2 remains enhanced regardless of the delay in the layer 1. Moreover, except for the same intra-layer delay, layer 1 always exhibits {a} suppressed chimera state as compared to the undelayed case with a small fluctuation in the critical coupling strength value. Next, our
study denotes the impact of parity of delay on the layer dynamics of the multiplex network. We introduce a small {value of} even delay in one layer (say layer 2), keeping another layer un-delayed. We find an immediate suppression of {the} chimera state in the delayed layer. Moreover, the chimera dynamics in the second layer remains suppressed regardless of the delay values in the first layer except {for} the small {value of} odd delay in layer 1 where it shows {an} enhancement. We find that as there is an increase in the delay {value} in both the layers, the suppression of chimera state against {the enhancement in} coupling strength becomes a dominant trait of the dynamics. Table~\ref{table.1} presents the critical coupling strengths for different delay values {for both the} layers of {a} multiplex network. The higher delay values in both the layers of {a} multiplex network lead to a suppressed chimera state even if there is a delay mismatch between the layers (Fig.~\ref{mul_4}). 

The chimera behavior for delayed dynamical evolution (Eq.~\ref{eq.evol}) can have a crucial dependence on the distribution of delays in the layers of the multiplex network. For a certain range of coupling strength, the appearance of chimera state can be destroyed and then again resurrected by controlling the parity and the distribution of delay in the dynamics of the multiplexed layers.
 
\vspace{3mm}
Furthermore, {it is noticed} that the inter-layer delay did not have any {significant} effect on the enhancement or suppression of the chimera state Fig.~\ref{mul_5}.{ An introduction of the homogeneous inter-layer delay ($\tau_{l1}=\tau_{l2}=0$ , $\tau_{12} \neq 0$), leads to all the nodes in both the layers experiencing the same delay in the information propagation across the layers and hence does not bring upon any significant change on the appearance of the chimera states.} We find that even with high inter-layer delay and no intra-layer delay, both {the} layers of the multiplex network exhibit exactly {the} same behavior with chimera state in the mid-range coupling values.
 \vspace{3mm}\\

\begin{table} [!tb]      
\caption{Critical coupling strengths for different delay values in the layers of multiplex network. upper(lower) triangles represents $\varepsilon_{critical}$ for layer 1 (layer 2). Other network parameters for each layer of the multiplex network are $N=100$ and $r=0.32$.}
\begin{center}
\begin{tabular} {|c|c|c|c|}
\hline 
\scriptsize \backslashbox{$\varepsilon^c_{l2}$}{$\varepsilon^c_{l1}$} & \scriptsize{$\tau_{(l1)}=0$} &\scriptsize {$\tau_{(l1)}=1$} &\scriptsize {$\tau_{(l1)}=2$} \\  \hline
\scriptsize {$\tau_{(l2)}=0$} &\scriptsize	 \backslashbox{0.49}{0.49}&\scriptsize	 \backslashbox{0.49}{0.52}&  \scriptsize \backslashbox{0.49}{0.41} \\ \hline
\scriptsize {$\tau_{(l2)}=1$}&\scriptsize	 \backslashbox{0.52}{0.49}&\scriptsize	 \backslashbox{0.53}{0.53}&  \scriptsize \backslashbox{0.52}{0.43} \\ \hline
\scriptsize {$\tau_{(l2)}=2$}&\scriptsize	 \backslashbox{0.41}{0.49}&\scriptsize	 \backslashbox{0.43}{0.52}&  \scriptsize \backslashbox{0.36}{0.36} \\ \hline 
\end{tabular}
\end{center}
\begin{flushleft} 
\label{table.1} 
\end{flushleft}
\end{table}
{\bf Conclusion.}
To summarize, we {have investigated} time-delayed dynamics for non-locally coupled chaotic maps and {have observed a} transition from {the} incoherent to {the} coherent dynamics via {the} chimera states. We demonstrate that the interplay of multiplexing and delay gives rise to {a} novel spatially clustered coherent states disconnected by incoherent regions known as chimera states. The emergence of chimera state in delayed systems shows a high dependency on the parity of the delay. Parity of delay is known to influence synchronization of the coupled dynamics \cite{delay_appl}. Here, we show that while the small odd (even) {value of} delay {leads to an} enhance{ment} (suppress{ion}) {in the} chimera state, {a} large delay {value leads to the suppression} regardless of the nature of the delay. {Our investigation has also uncovered the} layer chimera state with one coherent and one incoherent layer directly result{ing} from {the} enhancement-suppression behavior of individual layers of the multiplex network depending on the distribution of the delays. These results can provide additional insight into the formation of spatial clusters in delayed systems. Recently, similarities {between the emergence of chimera state in neural networks and EEG reading of a epileptic seizure state have been reported \cite{epileptic_2} with its possible applications in epileptic seizure diagnosis \cite{epileptic_1}}. Our result of {the} enhancement or {the} suppression of chimera state may {help in the diagnosis of} this kind of seizures by introducing a delay in the neural network. This finding may also contribute to enhancing our understanding of many biological functions known to show chimera-like states like uni-hemispheric sleep in humans and certain mammals.



{\bf Acknowledgments.} SJ  and SG thank DST project grant (EMR/2014/000368) and INSPIRE fellowship (IF150149), respectively. AZ acknowledges DFG support in the framework of SFB 910.

\end{document}